\documentclass[11pt,twoside]{article}

\usepackage{asp2006}
\usepackage{epsf}
\usepackage{psfig}
\usepackage{lscape}
\usepackage{graphicx}

\def\d3{$\delta_{3}$}
\def\1d3{$(1 + \delta_{3}) $}

\def\24m{24 $\mu$m}
\def\lir{$L_{IR}$ }

\def\sm{$M_{*}$ }
\def\kpc{$h^{-1}$kpc }

\def\speir{$L_{IR}/M_{*}$ }
\def\dis{$\vartriangle$$r_{p}$ }

\def\dis{$r_{p}$ }

\begin{document}
\title{Interaction-Triggered Star Formation in Distant Galaxies and the Role of Mergers in Galaxy Evolution}   
\author{Lihwai Lin}   
\affil{Institute of Astronomy \& Astrophysics,
Academia Sinica, Taiwan}    

\begin{abstract} 
The evolution of galaxy merger rates and its impact on galaxy properties have been studied intensively over the last decade. It becomes clear now that various types of mergers, i.e. gas-rich (wet), gas-poor (dry), or mixed mergers, affect the merger products in different ways. The epoch when each type of merger dominates also differs. In this talk, I review the recent progress on the measurements of galaxy merger rates out to $z \sim 3$ and the level of interaction-triggered star formation using large samples from various redshift surveys. These results provide insights to the importance of mergers in the mass assembly history of galaxies and in the evolution of galaxy properties. I also present new results in characterizing the environment of galaxy mergers, and discuss their implications in the built up of red-sequence galaxies.
\end{abstract}


\section{Introduction}   
Galaxy mergers have long been suggested to be associated with a variety of observational phenomena, including Ultra-Luminous Infrared Galaxies (ULIRGs), Quasars, post-starburst galaxies, Active Galactic Nuclei (AGN), and Submillimeter Galaxies (SMGs), etc. In certain galaxy evolution models, these objects may actually correspond to different phases during galaxy interactions \citep{hop06}. Galaxy mergers have significant impact on galaxy morphology, kinematics, stellar masses, and star formation histories of galaxies, as well as the mass of its central black holes. Recent studies on the galaxy luminosity function (LF) and stellar masses function (SMF) reveal that the number densities and stellar mass densities of galaxies in the red sequence have been increased by a factor of 2 to 3 since $z \sim 1$ while that of blue forming galaxies remain similar over the same period \citep{fab07,bel07}, suggesting a migration of galaxies from the blue cloud to the red sequence. One plausible mechanism responsible for this transformation is through merging of two blue galaxies, the so-called 'wet mergers' (gas-rich mergers) followed by subsequent mergers among re-sequence galaxies, the so-called 'dry mergers' (gas-poor mergers) \citep{fab07}. Therefore in order to better understand how the present-day massive galaxies are assembled, and how the galaxies have been evolved, we need to have improved constraints on the abundance of mergers and where they occur. There are three aspects regarding galaxy interactions that I focus on in this talk: the absolute galaxy merger rates, the influence of mergers in triggering star formation, and the environment of merging galaxies.

\section{Evolution of merger rates}
The are essentially two ways to identify interacting system. One is to pick up close pairs using the projected separation and redshift information. The advantage of using close pairs is that they are still well-separated and their properties have not been strongly influenced by the effect of mergers. This allows us to select interacting systems based on the properties of their progenitors such as their masses (or luminosities) and colors. The drawback is that these systems may not necessarily be close in physical space due to the difficulty in separating the Hubble flow from the peculiar velocity. It is therefore required to calibrate the true merger fraction among the galaxy pairs selected. The other approach relies on morphological classification. While visual classification of identifying mergers seem to be relatively robust, it becomes more challenging as going to higher redshifts and it is also very time-consuming for large surveys. What becomes more popular recently is to use non-parametric parameters such as CAS, denoted for concentration, asymmetry, and clumpiness, and Gini and M$_{20}$ measurements, where Gini is a measurement of flux distribution in galaxy's pixels, and M$_{20}$ is the 2nd order moment of light. It has been shown that the combinations of the above parameters can effectively pick up galaxies in the middle- and late-stage mergers \citep{con03,lot04,lot08}. However, it is not entirely clear which types of mergers (major vs minor mergers) are more sensitive to those measurements.

\begin{figure}
\centerline{\rotatebox{270}{\includegraphics[width=7.0cm]{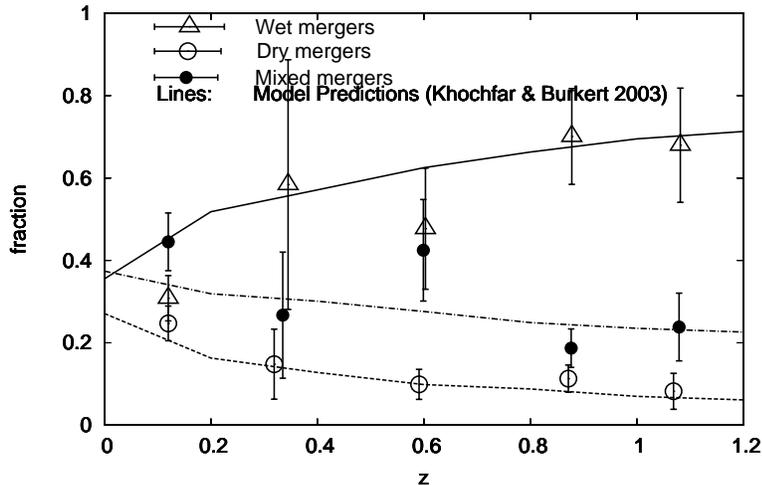}}}

\caption{Fraction of major mergers for wet (open triangles), dry (open circles), and
mixed mergers (solid circles) as a function of redshift. The symbols represent results from the DEEP2, TKRS, CNOC2
and MGC surveys. The three curves show the semi-analytical predictions of Sp-Sp (solid line), E-E (dashed line), and E-Sp (dash-dotted line) mergers by
\citet{kho03} but for a field-like environment (this figure is taken from Lin et al. 2008).  \label{figNmratio}}
\end{figure}

Using the aforementioned two approaches, there have been enormous progress made in order to study the time evolution of mergers rates. The evolution of merger fraction is conventionally parameterized by a power law of the form $(1+z)^{m}$. The exact values of $m$ have been diverse (see Lin et al. 2008 and references therein). This discrepancy comes from many factors, including how the merger samples are selected, what redshift ranges are probed, and what types of galaxies (luminosity, stellar mass, color, morphology, etc) are included in the samples. For example, it has been shown that the galaxy merger fraction depends strongly on galaxy luminosity and perhaps halo masses \citep{lin04,pat08,ste09,dom09}, and its redshift evolution is also dependent of the stellar mass \citep{con06}. Furthermore, the value of $m$ is a strong function of galaxy spectral types as well. A few studies have revealed that the frequency of dry mergers is not low and could be an important mechanism building up the masses of early-type galaxies \citep{van05,bel06}. In order to investigate the role of dry mergers in the evolution of galaxies and when it appears to become important, we classify candidates of wet, dry. and mixed mergers using colors of close pairs in the DEEP2 Redshift Survey. What we found is that blue galaxy pairs show slightly faster evolution with $m = 1.27\pm0.35$ while red-red pairs and red-blue pairs possess negative slope $m = -0.92\pm0.59$ and $-1.52\pm0.42$ respectively. Despite that the physical volume of the Universe increases with time as $(1+z)^{3}$, the departure of $m$ from 3 based on a simple density evolution argument is mainly due to the clustering effect of galaxies, as well as the change of blue and red galaxy populations. With certain assumptions on the merger time scale and the merger probability of kinematic pairs, we are able to investigate the relative roles of wet, dry, versus mixed mergers. As shown in Fig. \ref{figNmratio}, it is found that the relative fraction of wet mergers decreases with time, while those of dry  and mixed mergers keep growing since $z \sim 1$ \citep{lin08}. These results indicate that dry mergers play an important role at low redshift in assembling the mass of red-sequence galaxies, whereas wet mergers are being more essential at earlier epoch. Similar conclusions have also been drawn in the studies of VVDS data \citep{der09} and of GOODS data \citep{bun09}.

 Regardless of the difficulty in extending the merger rate measurement beyond $z \sim 1$ because of the need of deep near-infrared data, attempts to probe the merger rate up to $z \sim3$  have been made recently by several groups \citep{con03,rya08,blu09}. The exploration of merger rates at $z \sim 2-3$ is essential because this period is rough the peak of the cosmic star formation rate, quasar activities, and submillimeter sources. Based on a limit number of sample, it is suggested that the merger activities at $z \sim 2$ could be more frequent compared to $z \sim 1$ by a factor of 2-4. However, whether this trend continues to higher redshift, plateaus or peaks at $z \sim 2$ is still debatable. This will be a rich area for future studies that use HST/WFC3 and powerful ground-based near-infrared spectroscopy.

\section{Triggered star formation in interacting galaxies}
Hydrodynamic simulations of galaxy mergers that assume the density-dependent star formation predict that bursts of star formation occur after the first pass during the galaxy encounters \citep{mih96,bar04,cox06}. The idea that interactions between two gas-rich galaxies are effective in triggering star formation, however, is not new. Using local samples, Larson and Tinsley (1978) demonstrated that the distribution of interacting systems have bluer colors and a larger scatter in the (U - B, B - V ) diagram than normal galaxies, indicating a recent burst of star formation happening in the interacting galaxies. Another piece of evidence supporting the tidally-triggered star formation comes from the far-infrared selected samples: most ULIRGs (ultra-luminous infrared galaxies) in the local universe are found to be strongly interacting/merging systems (Borne 1999). However, ULIRGs are extreme populations of starbursting galaxies. The fact that they are mostly associated with mergers do not imply that galaxy interactions necessarily induce high-level of star formations. The key to constrain the strength of induced star formation in merging systems is to have large statistical samples of interacting galaxies and compare their star formation activities with those of isolated galaxies. Such exercises have been performed in studies from CfA2, 2dF, and SDSS surveys \citep{bar00,lam03,nik04}. These works found an anti-correlation between the star formation efficiency and kinematic parameters such as separations and velocity differences between pairs, supporting triggered star formation due to close encounters. It is worth noting that the level of SF enhancement seen in very close pairs is no greater than a factor of 2 compared to wide-separated pairs and control samples, even after restricting the samples to be late-type mergers. In other words, the efficiency of inducing star formation via galaxy interaction is not as dramatic as one would expect from studies on ULIRGs or HyperLIRGs.

\begin{figure}
\begin{center}
\includegraphics[angle=0,width=6.0cm]{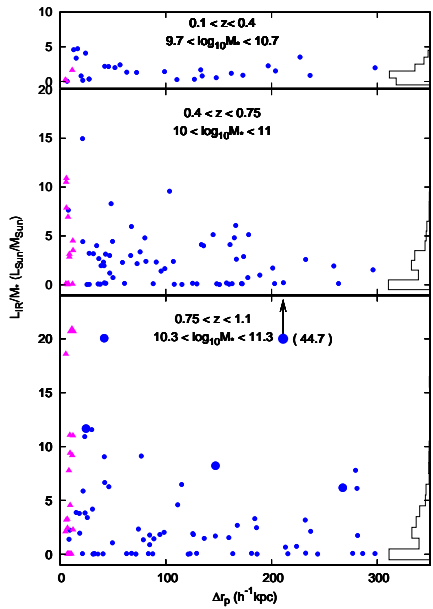}
\caption{The specific IR luminosity \speir as a function of
projected separation for kinematic galaxy pairs (circles) and
merging galaxies (triangles) in three redshift bins.
Merging galaxies are assigned to \dis $\sim5$ \kpc, which is the
minimum separation of kinematic close pairs. Galaxies with
$L_{IR}$ greater than $10^{12}$ $L_{\odot}$ (ULIRGs) are shown
with larger symbols. Distributions of \speir in control pairs are
also shown along the right-hand axes for comparison. Data points
at \speir = 0 refer to those sources with no \24m detection. The
number within the parenthesis in the bottom panel denotes the
\speir of that data point. Close pairs (\dis $<$ 50 \kpc) and
merging galaxies are found to have higher median \speir and wider
spread of \speir
than those of wide pairs (\dis $>$ 50 \kpc) and control pairs (see
Table 1) (taken from Lin et al. 2007). \label{fig2}}
\end{center}
\end{figure}

Now, what happens for galaxy interactions at higher redshift? There is consensus that the high-redshift galaxies are probably gas-richer, and their structures could be different from local samples (eg., they could be more bulgeless). If so, would that have significant impact on the activities during mergers? To explore this issue, we looked for interacting systems including close pairs and morphologically merging galaxies in the EGS field (one of the DEEP2 field), where we have Spitzer MIPS \24m observations \citep{lin07}. We measured the total infrared luminosity converted from \24m fluxes and the stellar masses for both interacting systems and field galaxies over the redshift range $0.1 < z < 1.1$. We find that while control samples form a nice sequence in the relation of \lir and \sm, close pairs and morphologically merging galaxies tend to lie on the upper end of this sequence. It is worth mentioning that \lir increases as we approach to higher redshifts for a given stellar mass range. At $z \sim 1$, the control sample can also exceed the ULIRG threshold without interactions as long as their stellar masses are large enough, in contrast to the case at very low redshifts, where mergers seems to be the only possible mechanism to boost the star formation up to the ULIRG level. Statistically we found that at $z\sim 1$, 7.1 $\pm$ 4.3\% of interacting galaxies with \sm $> 2\times10^{10}$ are found to be ULIRGs, compared to 2.6 $\pm$ 0.7\% in the control sample. If we plot the star formation rate efficiency versus the pair separation, again, we see an anti-correlation between these two parameters (Fig. \ref{fig2}). On average, we find the overall SFR enhancement during the interactions is close to a factor of 2. We do not find an evolution of this enhancement over the redshift range $0.1 < z < 1.1$.  However, we note that there is a large spread of SFR efficiency among interacting galaxies at a given separation, indicating that the starbursting efficiency also depend on other factors. It could be due to that we are sampling different stages of mergers, or due to the variation in internal properties of these interacting systems. If we interpret the spread of SFR seen in our sample is mostly attributed to differen merger stages, our results imply that the overall amount of enhanced star formation due to merger is probably too small to consume all the available gas in a couple of Gyrs.

There is still room for improvement of the above studies. In order to decipher whether it is the internal property of galaxies or the merger stage that dominate the scatter in the SFR of interacting galaxies, first it would be desirable to divide the samples based on various parameters, such as the gas contents, mass ratios, orbital configurations. Second is to have the ability to classify interacting systems into sequential merger stages. In addition to having larger samples, the development of techniques that can be used to map various stages of mergers will allow detailed comparisons with model predictions and also help to improve our understanding of mergers.

\section{Environment of mergers}
As discussed in \S 2, the dry mergers become more frequent at later times, mostly due to the increasing population of red galaxies toward low redshifts. On the other hand, we know that there exists well-known color-density relation \citep{hog04,coo06}, i.e., the fraction of red galaxies is greater in high-density regions compared to that in low-density regions. How the growth of red-sequence galaxies relates to the environment, and what is the role of mergers played in building up such color-density relation remains outstanding questions.

Another reason to study the environment of mergers is to separate the environment effect on intrinsic galaxy properties when we investigate the triggered star formation activities. \citet{bar07} demonstrates that most of close pairs reside in more massive halos compared to field galaxies. Given the fact that galaxy properties are tied to the environment, direct comparisons of the star formation rate between paired galaxies and isolated galaxies may be contaminated by the environment effect. Using constraints derived from cosmological N-body simulation, it is possible to apply isolation criterion for observed close pairs that are suitable for studies of tidally-triggered star formation.

\begin{figure}
\centerline{\rotatebox{90}{\includegraphics[width=7.5cm]{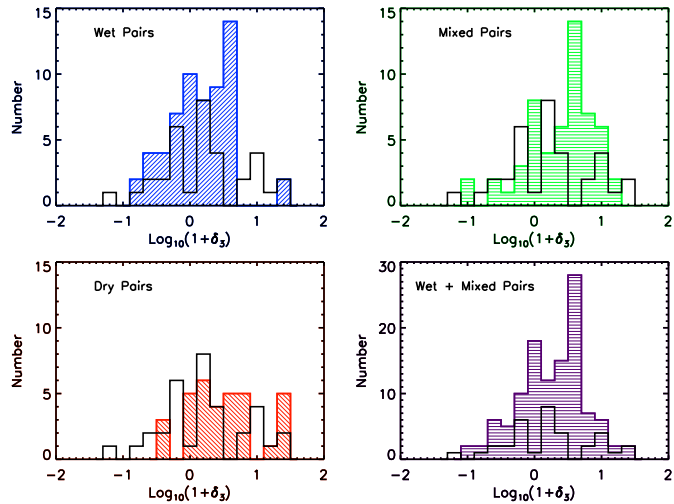}}}
\caption{The overdensity \1d3 distribution of wet (upper left panel), dry (lower-left panel), mixed (upper-right panel), and wet+mixed mergers (lower-right). The distributions of K+A galaxies are presented as solid curves while those of mergers are shown as shaded areas \citep{lin09}. \label{KA}}
\end{figure}

It has been recently shown that dry mergers preferentially occur in the groups and clusters. For example, \citet{tra08} suggested that dry mergers are important process to build up massive galaxies in the cores of galaxy groups/clusters by studying four X-ray luminous groups at intermediate redshifts ($z \sim 0.4$); \citet{mci08} also found that the frequency of mergers between luminous red galaxies (LRGs) is higher in groups and clusters compared to overall population of LRGs regardless of their environment by a factor of $2-9$ in the SDSS sample. Yet there lacks quantitative measurements of wet, dry, and mixed merger rates as a function of environments. One way to probe the connection between mergers and the formation of red galaxies is to compare the environment distribution of mergers to the post-starburst galaxies, the so-called 'K+A' or 'E+A' galaxies \citep{dre83}. These K+A galaxies are suggested to be the transition phase between star forming galaxies and the dead red-sequence galaxies, and hence they could be the direct progenitors of early-type red galaxies. In Fig. \ref{KA}, we show the overdensity distribution \1d3 \citep{coo06} of K+A galaxies found in DEEP2 sample \citep{yan09} versus wet, dry, and mixed mergers respectively, where \1d3 is the projected surface density relative to the mean density at a given redshift \citep{lin09}. It can be seen that the distributions of mixed and dry mergers alone are distinct from that of K+A galaxies, whereas wet mergers and wet plus mixed mergers distributes more similarly to K+A galaxies \citep{lin09}. This servers an indirect evidence about the linkage between mergers involving at least one gas-rich galaxy and the formation of K+A galaxies. Quantitative comparisons await bigger samples in both K+A galaxies and merger systems.

\section{Summary}
I have argued that studies of merger rates, effectiveness of the tidally-triggered star formation, and the host environment of mergers are the three keys to pin down the role of mergers in the evolution history of galaxies. Despite that our understanding of the above issues have been dramatically improved over the last few years, there remains several pieces of details missing in the whole picture. For example, under what kind of conditions will wet mergers lead to the quench of star formation and hence become K+A galaxies? What is the outcome of mixed mergers? What is the merger rate beyond redshift one? How well can we separate the contributions of minor mergers from major mergers? And finally, what is the role of AGN during galaxy encounters? Both detailed studies of individual interacting systems and statistical results based on future larger and deeper surveys of merging galaxies, together with improved numerical modeling that incorporate more sophisticated physics will provide essential knowledge to resolve the aforementioned issues.

\acknowledgements 
I thank the DEEP2 team and D. Patton for their contributions to several works presented here, E. Barton and D. McIntosh for their helpful discussions, and the conference organizers for a wonderful meeting.


\end{document}